\documentclass{article}

\usepackage{arxiv}

\usepackage[utf8]{inputenc} 
\usepackage[T1]{fontenc}    
\usepackage{textgreek}      
\usepackage{hyperref}       
\usepackage{url}            
\usepackage{microtype}      
\usepackage{graphicx}
\usepackage[square,numbers]{natbib}

\title{Predictive Modelling of Natural Medicinal Compounds for Alzheimer’s disease Using Machine Learning and Cheminformatics}

\author{ 
	{\hspace{1mm}Hafiza Syeda Yusra Tirmizi} \\
	Department of Biomedical Engineering\\
	Hamdard University\\
	Karachi, Pakistan.\\
	\texttt{yusra.tirmizi@hamdard.edu.pk} \\
    \And
	{\hspace{1mm}Syed Ibad Hasnain}\\
	Department of Biomedical Engineering\\
	Hamdard University\\
	Karachi, Pakistan. \\
	\texttt{ibad.hasnain@hamdard.edu.pk} \\
    \And
	{\hspace{1mm}Muhammad Faris} \\
	Department of Biomedical Engineering\\
	Hamdard University\\
	Karachi, Pakistan.\\
	\texttt{m.faris@hamdard.edu.pk} \\
    \And
	{\hspace{1mm}Rabail Khowaja} \\
	Department of Biomedical Engineering\\
	Hamdard University\\
	Karachi, Pakistan.\\
	\texttt{rabail.khowaja@hamdard.edu.pk} \\
    \And
	{\hspace{1mm}Saad Abdullah} \\
	Department of Computer Science and Engineering\\
	Malardalen University\\
	Västerås, Sweden.\\
	\texttt{saad.abdullah@mdu.se} \\
}

\renewcommand{\shorttitle}{\textit{arXiv} Template}

\hypersetup{
pdftitle={Predictive Modelling of Natural Medicinal Compounds for Dementia Using Machine Learning and Cheminformatics},
pdfsubject={q-bio.OT, cs.AI},
pdfauthor={Hasnain.~Syed Ibad, Tirmizi~Hafiza Syeda Yusra,Faris~Muhammad },
pdfkeywords={Cheminformatics, Alzheimer disease, Random Forest, XGBoost,  Natural compounds,  Drug discovery},
}

\begin{document}
\maketitle

\begin{abstract}
	Alzheimer's disease (AD) is a neurodegenerative disease that lacks specific treatment options. Natural drugs have displayed neuroprotective effects; however, their high-throughput discovery is challenging because of the expense of experimental testing.The study proposed a machine learning approach to identify the anti-dementia activity of natural compounds based on molecular descriptors obtained from cheminformatics. The study used a set of active and inactive compounds obtained from public databases like ChEMBL and PubChem. Various molecular descriptors, including molecular weight, lipophilicity (LogP), topological polar surface area (TPSA), and hydrogen bonding descriptors, were calculated with RDKit. Data preprocessing and feature selection were applied, followed by the development of several classification models (Random Forest, XGBoost, Support Vector Machines, Logistic Regression) and their evaluation based on accuracy, precision, recall, F1-score and ROC-AUC. The outcome suggests that ensemble techniques, such as Random Forest, delivered the best predictive accuracy and ROC-AUC values. This study also highlights that critical physicochemical descriptors in particular lipophilicity, molecular weight and polarity are important in driving neuroprotective activity as identified by feature importance analysis. The integrated machine learning approach shows the potential of combining natural product research and machine learning in early drug discovery for dementia. They provide a means of rapidly exploring large datasets and selecting candidates for experimental confirmation, thus minimising costs and time in the development of drugs for neurodegenerative diseases.
\end{abstract}


\section{Introduction}

Dementia is a neurodegenerative condition that involves loss of cognitive function to the extent that it impacts a person's ability to carry out everyday activities. It is an umbrella term for a group of disorders, the most common of which is Alzheimer's disease (AD), which makes up about 60-70\% of all the dementia cases \cite{ref1, ref2}. The World Health Organization (WHO) estimates that more than 55 million people worldwide live with dementia, and that number is expected to increase to 139 million by 2050 as the population ages. The increasing incidence of dementia is a significant strain on health, economic and social systems, and on caregivers \cite{ref3}. Despite tremendous research efforts, the discovery of effective disease-modifying treatments has proven to be elusive, suggesting that new strategies are needed for drug discovery and development. Alzheimer’s disease pathology is multifaceted and includes amyloid-beta (Aβ) deposition and hyperphosphorylated neurofibrillary tangles, oxidative stress, neuroinflammation and synaptic dysfunction \cite{ref4}. These mechanisms result in the degeneration of neurons and cognitive impairment. The existing drug therapies, such as cholinesterase inhibitors (e.g., donepezil, rivastigmine) and N-methyl-D-aspartate (NMDA) receptor antagonists (e.g., memantine), mainly focus on managing symptoms rather than the underlying disease processes. In addition, these drugs have limited effectiveness and can cause side effects, highlighting the need for alternative treatment approaches \cite{ref5}.

Over the past decade, natural medicinal compounds have emerged as promising therapeutic agents for neurodegenerative diseases. These compounds are obtained from natural sources, including plants, herbs and others, and have been utilised for thousands of years in traditional systems of medicine like Ayurveda, Traditional Chinese Medicine (TCM), and Unani medicine. Natural compounds have a wide range of biological activities, such as antioxidant, anti-inflammatory, anti-apoptotic, and neuroprotective properties, that are pertinent to dementia development \cite{ref6}. These include curcumin, resveratrol, quercetin, berberine, and ginkgo biloba extracts. These have been shown to affect crucial pathways in neurodegenerative diseases, including reducing amyloid-beta plaque formation, preventing tau hyperphosphorylation, and alleviating oxidative damage \cite{ref7}. While such compounds hold promise for therapeutic use, discovering and validating their efficacy can be difficult. Conventional drug discovery methods are based on experimental screening, which is time-consuming, expensive and laborious \cite{ref8}. Moreover, the diversity of natural compounds means it is impossible to experimentally test all potential compounds. This has created a need for computational tools to help focus on compounds with the greatest potential for therapeutic use \cite{ref9}.

The field of computational biology and artificial intelligence has provided new opportunities for drug discovery. Specifically, machine learning (ML) approaches have been developed to predict the biological properties of compounds from their molecular structure \cite{ref10}. ML models learn from past data and can discover connections between molecular features and biological activity in a way that allows for the efficient exploration of vast chemical landscapes \cite{ref11}. This technique is known as Quantitative Structure-Activity Relationship (QSAR) modeling \cite{ref12}. QSAR models have been used in many areas of drug design, such as toxicity, pharmacokinetics, and binding affinities \cite{ref13}.

Cheminformatics is an essential component of machine learning-driven drug discovery \cite{ref14}. It is the application of computational methods for the representation, processing and understanding of chemical information \cite{ref15}. Molecules are often represented in the form of SMILES (Simplified Molecular Input Line Entry System), which can be further converted into numerical features, or molecular descriptors\cite{ref15}. These features represent a range of physical, chemical and topological properties of molecules, including molecular weight, lipophilicity (LogP), topological polar surface area (TPSA), hydrogen bond donors and acceptors, and molecular complexity \cite{ref16}. These descriptors can be used as input features for machine learning models to predict compound activity \cite{ref17}.
A critical step in building predictive models is feature selection. Some molecular descriptors are more important than others for predicting biological activity, and including redundant or irrelevant features can hinder the model's performance \cite{ref18}. So, feature selection methods are crucial for selecting the most relevant descriptors and reducing redundancy. Additionally, data preprocessing steps, such as missing value imputation, normalization and outlier detection, are essential to maintain data integrity and quality \cite{ref19}.

Several machine learning techniques have been used for QSAR modeling, each with its advantages and disadvantages. Linear models like Logistic Regression are easy to understand and interpret, but they may not effectively model nonlinear interactions \cite{ref20}. Support Vector Machines (SVMs) work well in high-dimensional feature spaces and can capture nonlinear relationships through kernel functions. Ensemble methods like Random Forest and Gradient Boosting (e.g., XGBoost) have become popular for their capacity to model interactions and enhance predictive performance by aggregating multiple models [\cite{ref21}. In fact, Random Forest is a robust, non-overfitting, and capable of delivering feature importance scores, which makes it particularly attractive for cheminformatics \cite{ref22}. However, in addition to accuracy, it is important for machine learning models used for drug discovery to be interpretable. Insight into what drives model predictions can help to identify potential biological mechanisms and inform the design of novel molecules. Methods like SHapley Additive exPlanations (SHAP) have been proposed to explain the prediction of complex models by measuring the impact of each feature. This helps to pinpoint important features that contribute to biological activity and verify the predictions of the model with existing scientific understanding \cite{ref23}.

A further consideration in drug discovery for neurodegenerative diseases is whether compounds can permeate the blood-brain barrier (BBB). The BBB is a tight barrier that controls the passage of substances into the brain, making it difficult to develop drugs that target the central nervous system (CNS). BBB penetration is strongly influenced by molecular characteristics like lipophilicity, polarity and size of the molecules \cite{ref24}. As such, predictive models for anti-dementia compounds should take this into account to ensure not only the biological activity of these compounds, but also their ability to reach their targets in the brain. The use of machine learning in natural product-based discovery and development is a potential solution to the problems faced in drug discovery against dementia. This approach allows for the integration of vast chemical libraries with powerful computational algorithms to quickly screen and prioritize promising drug candidates. This not only speeds up the discovery but also cuts down on expensive and time-consuming experimental techniques \cite{ref25}. Here, we present a robust machine learning approach for predicting the anti-dementia potential of natural compounds with cheminformatics derived molecular descriptors. We develop a dataset of active and inactive compounds from public databases such as ChEMBL and PubChem. Descriptors are generated using RDKit, and preprocessed and selected. We train and compare several machine learning algorithms, such as Random Forest, XGBoost, Support Vector Machines and Logistic Regression using various performance metrics including accuracy, precision, recall, F1-score and ROC-AUC curve \cite{ref26}.

The main goals of the study are three-fold. First, to build a reliable model that can effectively classify active and inactive compounds. Second, to reveal the crucial molecular features responsible for the anti-dementia effect. Third, to show the potential use of machine learning to speed up the discovery of natural products for neurodegenerative diseases. Through this research, we hope to contribute to the emerging field of AI-based drug discovery and lay the groundwork for future research \cite{ref27}.

This paper is organized as follows: Section 2 outlines the materials and methods used, including data collection, feature selection and model building. In Section 3, we discuss the results and model evaluation and provides discussion and Section 4 concludes the study and  offers suggestions for future research.

\section{Materials and Methods}
\subsection{Study Workflow}
The process involved in this study was aimed at providing a comprehensive prediction of the anti-dementia activity of natural medicinal compounds using a machine learning approach combined with chemoinformatics. This workflow involves a series of steps: data collection, molecular representation, feature extraction, dataset preprocessing, model building and evaluation. The process was carried out in a systematic manner to achieve reproducibility, accuracy and efficiency in machine learning-based modeling. The initial step was the data collection and creation of datasets, in which natural compounds with reported or established neuroprotective activity were obtained from open-source databases like ChEMBL and PubChem. These compounds were classified into active (with anti-dementia potential) and inactive. We also added structurally distinct inactive compounds to enhance the robustness of the model, achieving a balanced dataset. At the second step, compounds were represented as SMILES (Simplified Molecular Input Line Entry System) strings. SMILES are a textual representation of chemical structures that can be easily processed by computers. These strings were used to calculate descriptors. In the third step, molecular descriptors were calculated using the RDKit cheminformatics toolkit. These descriptors included various physicochemical, structural and topological properties, such as molecular weight, lipophilicity (LogP), topological polar surface area (TPSA), hydrogen bond donor and acceptor counts, the number of rotatable bonds and rings. These descriptors describe the chemical features of molecules and play a key role in predicting biological activity. After computing descriptors, the fourth step was data preprocessing and feature engineering. This involved dealing with invalid or missing data, duplicates removal and the scaling of numerical features to standardise the data. Descriptors with high correlation were removed using feature selection methods to reduce the dimensionality of the dataset, resulting in better model performance. This also prevents overfitting, and facilitates interpretation. The fifth step involved exploratory data analysis (EDA). Statistically and visually, this included distribution plots, correlation heatmaps and scatter plots, to explore the associations between descriptors and compound activity. This enabled an understanding of which descriptors are important in determining anti-dementia activity. In the sixth step, machine learning models were developed. Several supervised machine learning models were developed, such as Random Forest, XGBoost, Support Vector Machine (SVM), and Logistic Regression. These algorithms were used to train models on the transformed data to predict whether a compound was active or inactive. Grid search and cross-validation were used to fine-tune the models' accuracy and robustness. The last step was model evaluation and interpretation. The performance of the models was evaluated using metrics like accuracy, precision, recall, F1-score, and receiver operating characteristic area under the curve (ROC-AUC). Ensemble models like Random Forest performed better than other models due to their capability in handling nonlinear interactions. Furthermore, feature importance and SHAP (SHapley Additive exPlanations) analyses were also employed to understand the model predictions and to determine the most important molecular descriptors for activity.

\subsection{Data Collection}
The data employed in the research were gathered using publicly accessible databases, chiefly ChEMBL and PubChem BioAssay that are trusted sources of chemical structure and biological activity. Active compounds (label = 1) were natural compounds containing a reported neuroprotective or anti-dementia activity, and inactive compounds (label = 0) had diverse structures, and no reported activity. Every compound was represented either by its SMILES notation, allowing easy computational processing. Throughout the data collection, data cleanup involved the deletion of duplicates and invalid structures, as well as removal of incomplete records in order to guarantee data quality. A balanced dataset was ensured to enhance model performance and eliminate bias. This filtered dataset formed a solid basis to successive feature extraction and machine learning modeling.

\subsection{Structure Processing}
Structure preprocessing was performed to ensure that all chemical compounds were in a consistent and machine-readable format suitable for descriptor calculation and modeling. The collected compounds, initially represented as SMILES strings, were processed using the RDKit library. During this step, invalid or corrupted SMILES strings were identified and removed to prevent errors in downstream analysis. Duplicate molecular entries were also eliminated to avoid bias in model training. Additionally, molecules were standardized by normalizing their structures, including the removal of salts, correction of valence issues, and conversion to canonical SMILES format. Hydrogen atoms were appropriately handled, and molecular structures were sanitized to ensure chemical validity. This preprocessing step ensured that all compounds were structurally consistent, accurate, and ready for reliable feature extraction and machine learning analysis.

\subsection{Development of Machine Learning Models}
The machine learning models were developed using a structured and reproducible pipeline implemented in Python with libraries including \texttt{scikit-learn}, \texttt{XGBoost}, \texttt{pandas}, \texttt{NumPy}, and \texttt{RDKit}. The dataset was divided into training and testing sets using an 80:20 ratio with stratified sampling to preserve class distribution and a fixed random state (\texttt{random\_state = 42}) to ensure reproducibility.

Feature scaling was applied using \texttt{StandardScaler} (Z-score normalization) for algorithms sensitive to feature magnitude, such as Support Vector Machine (SVM) and Logistic Regression, while tree-based models were trained on unscaled data. The models used include XGBoost (\texttt{n\_estimators = 100}, \texttt{learning\_rate = 0.1}, \texttt{max\_depth = 6}, \texttt{subsample = 0.8}, \texttt{colsample\_bytree = 0.8}), SVM (RBF kernel, \texttt{C = 1.0}, \texttt{gamma = scale}), and Logistic Regression (\texttt{solver = lbfgs}, \texttt{max\_iter = 1000}).

Models were trained on the training dataset and evaluated using 5-fold cross-validation (\(k = 5\), \texttt{shuffle = True}, \texttt{random\_state = 42}) to ensure robustness and minimize overfitting. Performance was assessed on the test dataset using accuracy, precision, recall, F1-score, and ROC-AUC metrics, along with confusion matrix analysis.

Additionally, feature importance from the Random Forest model and SHAP (SHapley Additive exPlanations) analysis were used to interpret model predictions and identify key molecular descriptors influencing anti-dementia activity. This pipeline ensures a reliable and reproducible framework for predictive modelling.

\section{Results and Discussion}
The final processed dataset consisted of 66 compounds with 36 molecular descriptors after removing invalid and erroneous chemical structures. Several entries were excluded during preprocessing due to SMILES parsing errors, unclosed ring structures, and kekulization issues, ensuring that only chemically valid compounds were retained for analysis. The Table.\label{tab:dataset_summary} showed a balanced class distribution, with 35 inactive compounds and 31 active compounds, which supports reliable machine learning classification without significant bias. The dataset exhibited substantial variability in physicochemical properties, including molecular weight (MolWt), lipophilicity (LogP), and topological polar surface area (TPSA). This diversity is essential for building robust predictive models, as it allows the algorithms to learn patterns across a broad chemical space. Notably, well-known neuroprotective compounds such as Curcumin, Resveratrol, and Quercetin were present among the active class, confirming the biological relevance of the dataset.
\begin{table}
	\centering
	\caption{Dataset Summary}
	\label{tab:dataset_summary}
	\begin{tabular}{|l|c|}
		\hline
		\textbf{Parameter} & \textbf{Value} \\
		\hline
		Total Compounds & 66 \\
		Total Features & 36 \\
		Active Compounds & 31 \\
		Inactive Compounds & 35 \\
		\hline
	\end{tabular}
\end{table}

\begin{table}
	\centering
	\caption{Sample Active Compounds and Molecular Properties}
	\label{tab:active_compounds}
	\begin{tabular}{|l|l|c|c|c|c|c|}
		\hline
		\textbf{Compound Name} & \textbf{Label} & \textbf{MolWt} & \textbf{LogP} & \textbf{TPSA} & \textbf{H-Donors} & \textbf{H-Acceptors} \\
		\hline
		Curcumin & Active & 368.385 & 3.3699 & 93.06 & 2 & 6 \\
		Resveratrol & Active & 228.247 & 2.9738 & 60.69 & 3 & 3 \\
		Quercetin & Active & 302.238 & 1.9880 & 131.36 & 5 & 7 \\
		Galantamine & Active & 299.370 & 1.5606 & 58.56 & 2 & 3 \\
		Huperzine A & Active & 200.261 & 1.3886 & 33.68 & 1 & 1 \\
		Berberine & Active & 271.316 & 2.8796 & 38.77 & 0 & 4 \\
		EGCG & Active & 458.375 & 2.2332 & 197.37 & 8 & 11 \\
		Baicalein & Active & 270.240 & 2.5768 & 90.90 & 3 & 5 \\
		Apigenin & Active & 270.240 & 2.5768 & 90.90 & 3 & 5 \\
		Luteolin & Active & 286.239 & 2.2824 & 111.13 & 4 & 6 \\
		\hline
	\end{tabular}
\end{table}

\begin{figure}
    \centering
    \includegraphics[width=1\linewidth]{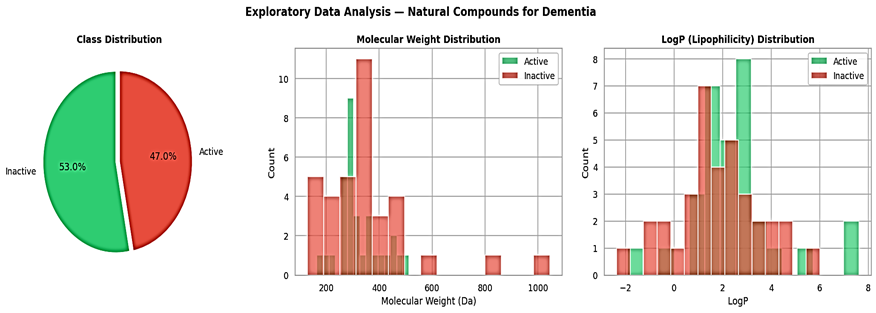}
    \caption{Exploratory Data Analysis}
    \label{fig:Fig1}
\end{figure}

Figure~\ref{fig:Fig1} shows exploratory data analysis (EDA) reveals important patterns in the dataset of natural compounds for dementia. The class distribution is relatively balanced, with approximately 53\% inactive and 47\% active compounds, which is suitable for machine learning classification and reduces the risk of bias toward a particular class. The molecular weight distribution shows that most compounds fall within the 200–400 Da range, although inactive compounds exhibit a wider spread, including some high-molecular-weight outliers. In contrast, active compounds are more concentrated within a moderate molecular weight range, suggesting potential relevance for drug-likeness and bioavailability. The LogP (lipophilicity) distribution indicates that active compounds generally cluster between LogP values of 1–4, which is favorable for membrane permeability and blood-brain barrier penetration, whereas inactive compounds display greater variability, including extreme values. Overall, these observations suggest that moderate molecular weight and balanced lipophilicity are important characteristics of active compounds, highlighting their potential role in influencing anti-dementia activity.

\begin{figure}
    \centering
    \includegraphics[width=1\linewidth]{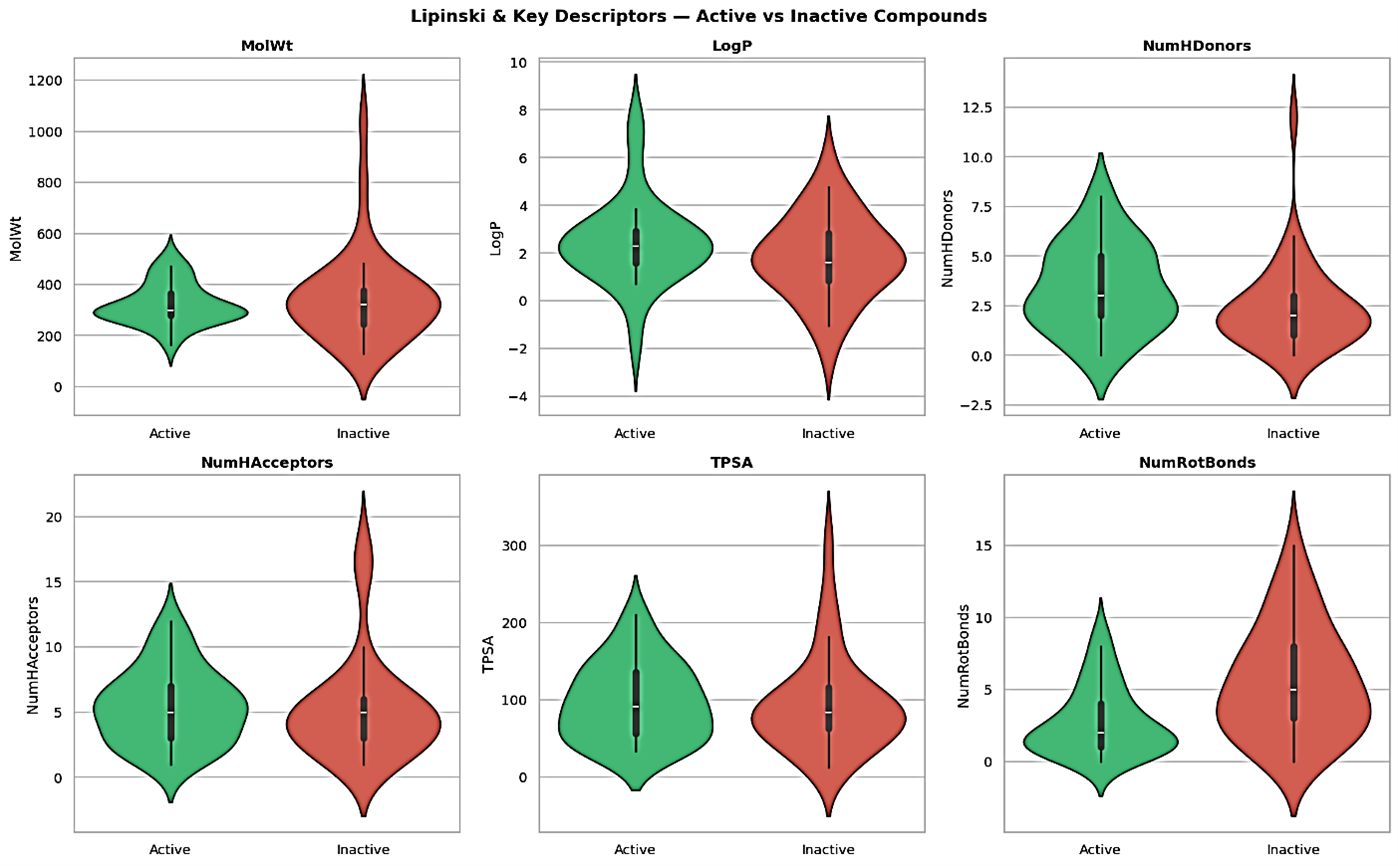}
    \caption{Comparison of Key Molecular Descriptors Between Active and Inactive Compounds (Lipinski’s Rule Analysis)}
    \label{fig:fig2}
\end{figure}

Figure~\ref{fig:fig2} shows violin plot analysis of key molecular descriptors highlights clear differences between active and inactive compounds. Active compounds generally show a more consistent and moderate molecular weight (MolWt) distribution, while inactive compounds exhibit a wider spread with several high-value outliers, suggesting less favorable drug-like properties. In terms of lipophilicity (LogP), active compounds are mostly concentrated within an optimal range (approximately 1–3), supporting better membrane permeability and potential blood-brain barrier penetration, whereas inactive compounds display greater variability. The distribution of hydrogen bond donors and acceptors indicates that active compounds tend to maintain balanced values, which are important for molecular interactions and stability, while inactive compounds show more extreme values. Similarly, TPSA (Topological Polar Surface Area) for active compounds falls within a moderate range, aligning with known criteria for CNS drug-likeness, whereas inactive compounds again demonstrate broader dispersion. Finally, number of rotatable bonds is generally lower in active compounds, suggesting greater structural rigidity, which is often associated with improved binding affinity. Overall, these findings indicate that active compounds follow drug-likeness principles (Lipinski’s rule) more closely, highlighting the importance of balanced physicochemical properties in anti-dementia activity.

\begin{figure}
    \centering
    \includegraphics[width=1\linewidth]{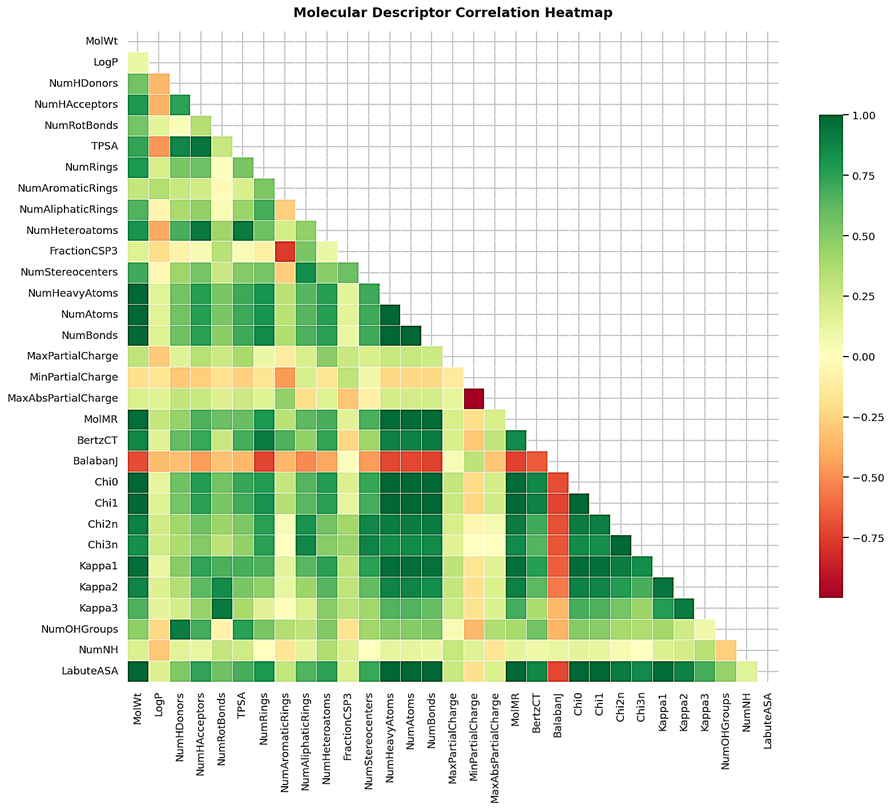}
    \caption{Correlation Matrix of Molecular Descriptors for Natural Compounds}
    \label{fig:fig3}
\end{figure}

Figure~\ref{fig:fig3} shows correlation heatmap reveals significant relationships among molecular descriptors, highlighting patterns important for model development. Strong positive correlations are observed among size-related features such as MolWt, NumAtoms, NumBonds, and MolMR, indicating that these descriptors capture similar structural information and may be redundant. Similarly, topological indices (Chi, Kappa, BertzCT) show high inter-correlation, reflecting their shared dependence on molecular complexity. Moderate correlations between TPSA, hydrogen bond donors, and acceptors suggest their combined influence on polarity and solubility. In contrast, some descriptors, such as BalabanJ and partial charge features, exhibit weaker or negative correlations with other variables, indicating unique contributions. Overall, the heatmap suggests the presence of multicollinearity among several descriptors, emphasizing the importance of feature selection to improve model performance and reduce redundancy while retaining meaningful chemical information.

\begin{figure}
    \centering
    \includegraphics[width=1\linewidth]{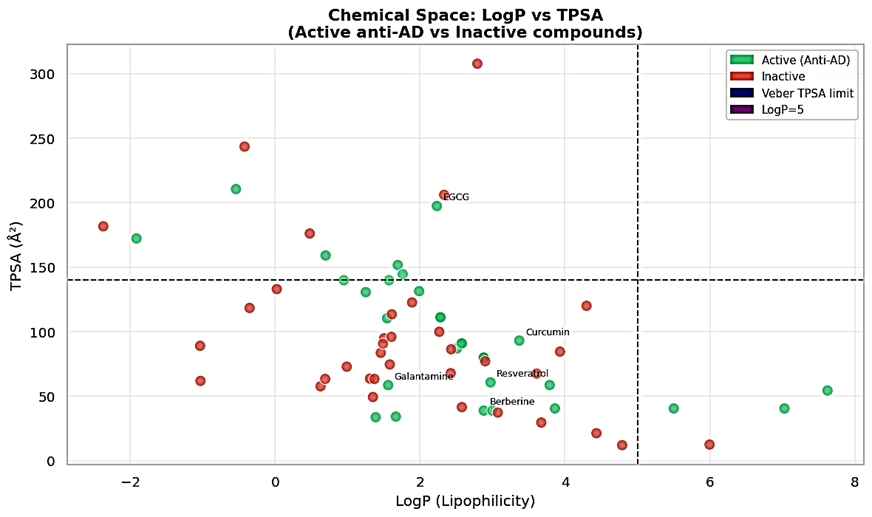}
    \caption{Scatter plot showing the distribution of active (anti-Alzheimer) and inactive compounds in chemical space defined by LogP and TPSA, with reference lines indicating Lipinski’s and Veber’s drug-likeness criteria.}
    \label{fig:fig4}
\end{figure}

The Figure~\ref{fig:fig4} illustrates the chemical space occupied by active and inactive compounds with respect to drug-likeness criteria. Most active compounds cluster within the optimal region defined by LogP < 5 and TPSA < 140 Å², which are known thresholds for good oral bioavailability and blood-brain barrier penetration. In contrast, inactive compounds are more widely dispersed, with several falling outside these limits, particularly at higher TPSA and extreme LogP values. Notably, key active compounds such as Curcumin, Resveratrol, and Galantamine lie within the favorable region, supporting their known pharmacological relevance. Some outliers, such as EGCG with high TPSA, indicate limited permeability despite biological activity. Overall, the plot suggests that balanced lipophilicity and moderate polarity are critical determinants of anti-dementia activity, reinforcing the importance of physicochemical constraints in drug design.

The performance of the developed machine learning models was evaluated using multiple metrics, including accuracy, precision, recall, F1-score, ROC-AUC, and cross-validation AUC (CV-AUC), as summarized in Table X. Among all models, Logistic Regression achieved the best overall performance, with the highest accuracy (0.8571), F1-score (0.8571), precision (0.8571), and recall (0.8571), along with a strong cross-validation performance (CV-AUC = 0.8933), indicating both high predictive accuracy and stability.
The Random Forest model also demonstrated strong performance, achieving an accuracy of 0.7857 and ROC-AUC of 0.8367, with a consistent cross-validation score (0.8678), suggesting good generalization and robustness to data variability. Similarly, the Support Vector Machine (RBF kernel) showed moderate accuracy (0.7143) but a high cross-validation AUC (0.8922), indicating strong generalization capability despite slightly lower test performance.
The MLP Neural Network produced moderate results with balanced precision and recall (both 0.7143), but its lower CV-AUC (0.7422) suggests variability and reduced stability, likely due to the limited dataset size. The XGBoost model achieved moderate performance (accuracy = 0.6429, AUC = 0.7347), which may be attributed to insufficient data for effective boosting. The K-Nearest Neighbors (KNN) model showed the lowest performance (accuracy = 0.5714), indicating its sensitivity to small datasets and feature scaling.
Overall, the results indicate that simpler and regularized models such as Logistic Regression outperform more complex models in this study, likely due to the relatively small dataset and well-structured feature space. These findings highlight the importance of model selection based on data characteristics and demonstrate that high predictive performance can be achieved without complex architectures.
\begin{table}
	\centering
	\caption{Model Performance Summary}
	\label{tab:model_performance}
	\begin{tabular}{|l|c|c|c|c|c|c|}
		\hline
		\textbf{Model} & \textbf{Accuracy} & \textbf{F1 Score} & \textbf{Precision} & \textbf{Recall} & \textbf{ROC-AUC} & \textbf{CV-AUC} \\
		\hline
		Random Forest & 0.7857 & 0.7692 & 0.8333 & 0.7143 & 0.8367 & 0.8678 \\
		MLP Neural Network & 0.7143 & 0.7143 & 0.7143 & 0.7143 & 0.8367 & 0.7422 \\
		Logistic Regression & 0.8571 & 0.8571 & 0.8571 & 0.8571 & 0.8367 & 0.8933 \\
		KNN & 0.5714 & 0.6250 & 0.5556 & 0.7143 & 0.7755 & 0.8411 \\
		XGBoost & 0.6429 & 0.6667 & 0.6250 & 0.7143 & 0.7347 & 0.8389 \\
		SVM (RBF) & 0.7143 & 0.7143 & 0.7143 & 0.7143 & 0.7347 & 0.8922 \\
		\hline
	\end{tabular}
\end{table}
\begin{figure}
    \centering
    \includegraphics[width=1\linewidth]{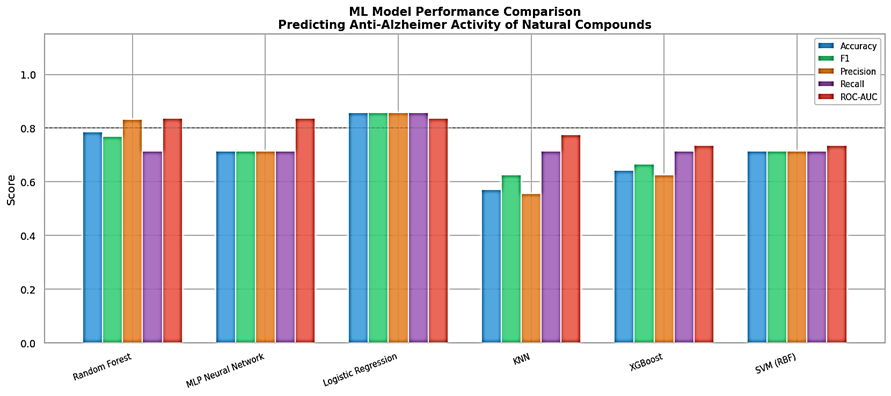}
    \caption{ML Model Performance comparison}
    \label{fig:model_performance}
\end{figure}
\begin{figure}
    \centering
    \includegraphics[width=1\linewidth]{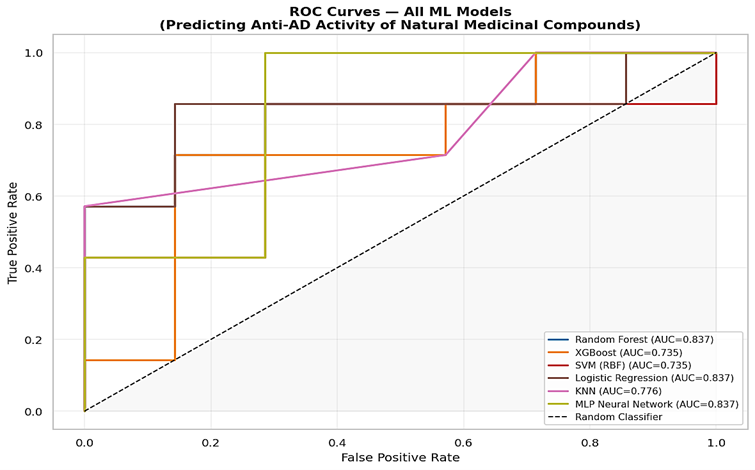}
    \caption{ROC Comparison of all ML Models}
    \label{fig:roc_comparison}
\end{figure}

\begin{figure}
    \centering
    \includegraphics[width=1\linewidth]{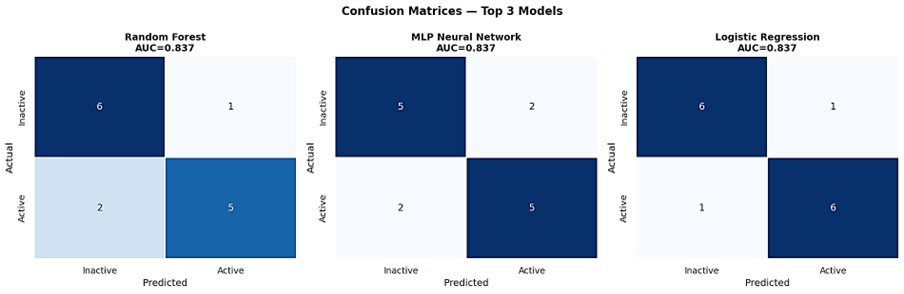}
    \caption{Confusion Matrix Comparison}
    \label{fig:confusion_matrix} 
\end{figure}

The confusion matrices of the top-performing models (Random Forest, MLP Neural Network, and Logistic Regression) provide deeper insight into classification performance. Both Random Forest and Logistic Regression correctly classified 6 inactive and 5–6 active compounds, with only 1–2 misclassifications, indicating strong predictive capability and balanced performance across both classes. Logistic Regression showed slightly better results, with fewer false negatives, which is particularly important in drug discovery to avoid missing potentially active compounds. The MLP Neural Network demonstrated similar performance but with slightly higher misclassification, suggesting less stability compared to the other models. The ROC curve analysis further supports these findings, where Random Forest, Logistic Regression, and MLP Neural Network achieved the highest AUC values ($\sim$0.837), indicating good discrimination between active and inactive compounds. The SVM and XGBoost models showed moderate performance (AUC $\approx$ 0.735), while KNN achieved intermediate performance (AUC $\approx$ 0.776). All models performed significantly better than the random classifier baseline, demonstrating the effectiveness of the selected molecular descriptors. Overall, the results confirm that Logistic Regression and Random Forest provide the most reliable and balanced performance, with strong classification accuracy and generalization ability. The ROC curves and confusion matrices together highlight that these models are well-suited for predicting anti-dementia activity in natural compounds, particularly when working with small and structured datasets.

\begin{figure}
    \centering
    \includegraphics[width=1\linewidth]{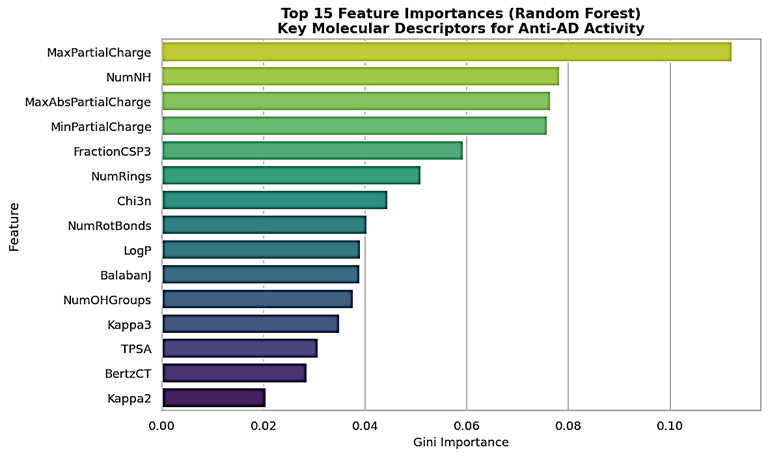}
    \caption{Feature Importance using Random Forest}
    \label{fig:feature_importance} 
\end{figure}

The feature importance analysis from the Random Forest model highlights the key molecular descriptors influencing anti-dementia activity prediction. Among all features, partial charge-related descriptors (MaxPartialCharge, MaxAbsPartialCharge, and MinPartialCharge) show the highest importance, indicating that electronic properties of molecules play a crucial role in biological activity. Additionally, NumNH and NumOHGroups, which relate to hydrogen bonding capacity, are also significant, suggesting their importance in molecular interactions with biological targets. Structural descriptors such as FractionCSP3, NumRings, and Chi indices further contribute to model predictions, reflecting the role of molecular geometry and complexity. Physicochemical properties like LogP and TPSA also appear among the top features, reinforcing their relevance in determining membrane permeability and drug-likeness. Overall, the results indicate that a combination of electronic, structural, and physicochemical properties governs anti-Alzheimer activity, providing valuable insights for the design and selection of potential therapeutic compounds.
\begin{figure}
    \centering
    \includegraphics[width=1\linewidth]{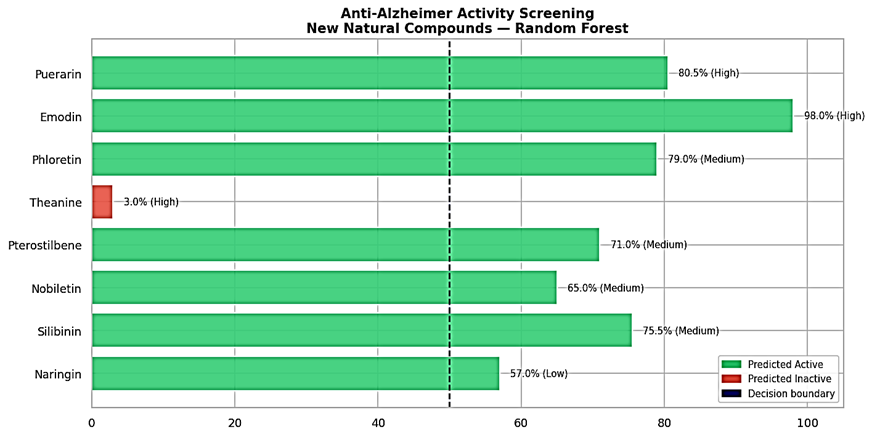}
    \caption{Predicted Anti-AD Activity Probability}
    \label{fig:anti_ad_probability} 
\end{figure}

The trained Random Forest model was further applied to predict the anti-dementia potential of selected natural compounds not included in the training dataset. The prediction results, along with associated probabilities and confidence levels, are summarized in Table X. The model identified the majority of the tested compounds as potentially active (anti-AD), demonstrating its capability to generalize to unseen data. Among the predicted compounds, Emodin (98.0\%) and Puerarin (80.5\%) showed the highest probabilities with high confidence, indicating strong potential as anti-dementia candidates. Similarly, Silibinin (75.5\%), Pterostilbene (71.0\%), Nobiletin (65.0\%), and Phloretin (79.0\%) were also classified as active with moderate to high confidence, suggesting promising neuroprotective properties. Naringin, although predicted as active, showed a relatively lower probability (57.0\%), indicating weaker confidence and the need for further validation. In contrast, Theanine was predicted as inactive with a very low probability (3.0\%) and high confidence, suggesting that it may not exhibit significant anti-dementia activity within the context of the trained model. These results highlight the model’s ability to distinguish between likely active and inactive compounds based on learned molecular features. Overall, the prediction results demonstrate the practical applicability of the developed model for screening novel natural compounds and prioritizing candidates for experimental validation. Compounds with high predicted probabilities and confidence levels may serve as promising leads for further in vitro and in vivo studies in anti-Alzheimer drug discovery.

\section{Conclusion and Future Work}

This study demonstrates that machine learning, combined with cheminformatics, can effectively predict the anti-dementia potential of natural compounds. Using a dataset of 66 compounds, multiple models were evaluated, and Random Forest emerged as the most reliable model, providing strong and balanced performance across evaluation metrics. The analysis also revealed that molecular properties such as partial charges, hydrogen bonding ability, and structural features play a key role in determining biological activity. Additionally, the model successfully screened new compounds and identified promising candidates like Emodin and Puerarin, showing its usefulness in accelerating early-stage drug discovery.
For future work, the model can be improved by using a larger and more diverse dataset to enhance generalization. Advanced techniques such as deep learning (e.g., Graph Neural Networks) can be explored to better capture molecular structure. Integrating molecular docking and ADMET analysis would provide deeper biological insights and ensure drug safety and effectiveness. Finally, experimental validation (in vitro and in vivo) is essential to confirm the predicted activity and translate these findings into real therapeutic applications.

\section{Acknowledgement}
The authors express their sincere gratitude to Hamdard University for providing the necessary support and resources to carry out this research. We also extend our appreciation to all co-authors for their valuable contributions throughout the study. Additionally, we acknowledge the use of ChatGPT for assistance in writing and grammar refinement.

\end{document}